\documentstyle[prl,aps]{revtex}
\begin{document}

\draft
\title{The Effect of Focusing and Caustics \\on Exit Phenomena
in Systems Lacking
Detailed Balance}
\author{Robert S. Maier${}^{(1)}$ and D.~L. Stein${}^{(2)}$}
\address{Mathematics${}^{(1)}$ and Physics${}^{(2)}$ Departments,
University of
Arizona, Tucson, Arizona 85721, U.S.A.}

\maketitle
\begin{abstract}
We study the trajectories followed by a particle subjected to weak noise
when escaping from the domain of attraction of a stable fixed point.
If~detailed balance is absent, a~{\em focus\/} may occur along the most
probable exit path, leading to a breakdown of symmetry (if~present).  The
exit trajectory bifurcates, and the exit location distribution may become
`skewed' (non-Gaussian).  The weak-noise asymptotics of the mean escape
time are strongly affected.  Our methods extend to the study of skewed exit
location distributions in stochastic models without symmetry.
\end{abstract}

\narrowtext

A~particle moving in a force field, but weakly perturbed by external noise,
will spend most of its time near stable fixed points of the force field.
But the particle will occasionally undergo a {\em large fluctuation\/}:
it~will leave the basin of attraction of one such point, and enter that of
another.  The timescale on which such fluctuations occur grows
exponentially in the low-noise limit.

In~this limit, the particle typically follows a {\em unique trajectory\/}
in its final, successful escape attempt; transverse deviations from this
path become increasingly unlikely as the noise strength goes to zero.  This
trajectory is referred to as the most probable escape path (MPEP), and its
properties govern the asymptotic behavior of the mean escape time.  In~our
earlier work~\cite{MS1,MS2} we~suggested the possibility of a {\em focusing
singularity\/}: as~one moves out along the MPEP, transverse fluctuations
become increasingly less stable, leading to a breakdown of stability before
the boundary of the basin is reached.  This hitherto unexplored phenomenon
has profound consequences for the particle's escape behaviour.
It~frequently occurs when the force field is nonconservative, or in~general
when detailed balance is absent.  Its~wide prevalence has not previously
been recognized.

In this Letter we study a simple example to demonstrate the effects of
focusing singularities on exit phenomena.  We~show that when such a
singularity occurs, the MPEP {\em bifurcates\/} (in~the simplest case),
leading to a type of broken symmetry.  This transition affects not only the
prefactor of the mean first passage time (MFPT) to the boundary, but also,
its {\em exponential dependence on the noise\/}.  It~also induces an
unusual `skewed' (non-Gaussian) distribution of exit points on the
boundary.

For simplicity we restrict ourselves to the case of an overdamped classical
point particle moving in two dimensions and subject to additive isotropic
white noise~${\bf w}(t)$.  In~the overdamped case the deterministic forces
acting on the particle are described by a drift field~${\bf u}({\bf x})$, so
that
\begin{equation}
dx_i(t) = u_i({\bf x}(t))\,dt + \epsilon^{1/2}\,dw_i(t),\quad i=1,2
\end{equation}
with $\epsilon$ the noise strength.  The corresponding Fokker-Planck
equation for the probability density is
\begin{equation}
\dot\rho = (\epsilon/2)\nabla^2\rho - {\bf\nabla} \cdot (\rho {\bf u}).
\end{equation}
We~shall employ the technique of matched asymptotic expansions described in
Refs.~\cite{MS1,MS2}.  We~impose absorbing boundary conditions on the
`separatrix' (the boundary of the basin) and study the behavior,
as~$\epsilon\to 0$, of the quasistationary density~$\rho_1(x,y)$,
{\em i.e.}, the
slowest decaying eigenmode of the Fokker-Planck operator.  In~the low-noise
limit, the properties of~$\rho_1$ can be used to calculate both the MFPT
and the distribution of exit points~\cite{MS1}.

We~specialize to the case of a drift field with a stable fixed point
at~$(x_S,0)$, whose basin of attraction is the entire right-half plane.
We~assume symmetry about the $x$-axis, and that there is only one fixed
point on the $y$-axis: a~saddle point at~$(0,0)$.
A simple drift field with this structure is
\begin{equation}
\label{eq:drift}
{\bf u}(x,y) =(x-x^3-\alpha xy^2,{}-\mu y - x^2y).
\end{equation}
This force field is
conservative, {\em i.e.}, is~the gradient of a potential, only
if~$\alpha=1$.  Nonetheless for any $\mu>0$ the field conforms to the
assumptions, with $x_S=1$.  The symmetry of~${\bf u}$ suggests that the
MPEP, which must connect $(x_S,0)$ and the saddle, lies along the $x$-axis,
and for small~$\alpha>0$ that is the case~\cite{MS2}.  The case when exit
occurs over an~{\em unstable\/} fixed point~\cite{MS1} is also of interest;
additional phenomena may emerge there because focusing on the separatrix
(rather than the~MPEP) can occur.  We~defer that case to a later paper.

Away from both the stable point and the separatrix, $\rho_1$~can be
approximated by a WKB form~\cite{MS2,Talkner87}
\begin{equation}
\label{eq:WKB}
\rho_1(x,y)\sim K(x,y)\exp \left( -W(x,y)/\epsilon\right).
\end{equation}
$K(x,y)$~satisfies a transport equation, and $W(x,y)$ an eikonal
(Hamilton-Jacobi) equation: $H({\bf x},{\bf\nabla} W)=0$, with $H({\bf
x},{\bf p})={\textstyle{1\over2}}{{\bf p}}^2+{\bf u}({\bf x})\cdot {\bf p}$ the
Wentzell-Freidlin Hamiltonian~\cite{VF}.  So~for any point~$(x,y)$ in the
basin of attraction of~$(x_S,0)$, $W(x,y)$~will be the action of a
zero-energy classical trajectory, governed by this Hamiltonian, extending from
$(x_S,0)$ to~$(x,y)$.  In~general, computation of~$W(x,y)$ will
require a minimization over the set of zero-energy trajectories terminating
at~$(x,y)$.  MPEPs are accordingly the zero-energy trajectories from
$(x_S,0)$ to~$(0,0)$ of~least action.  The MPEP action is the exponential
growth rate, in~the low-noise limit, of the MFPT.

For non-gradient drift fields~${\bf u}$, it~will be a common occurrence
for two or more classical trajectories emanating from~$(x_S,0)$ to~cross at
some point.  The locus of these points (equivalently, the set of points
reachable from~$(x_S,0)$ {\em vi\^a\/} more than one zero-energy
trajectory) is known as a
{\em caustic\/}~\cite{Gutzwiller90,Schulman,Chinarov93}.
A~{\em focus\/} is a critical point from which a caustic emanates.
Pictorial examples of this are given in
Refs.~\cite{MS2,Chinarov93,Day87}.
In~these examples the caustics are found far from fixed points, MPEPs, or
separatrices and have little effect on exit phenomena.  However they may
constitute regions of anomalous probability density, since the WKB
expansion breaks down there~\cite{Schulman}.

Our concern here is what happens when a focus appears on the MPEP itself,
making the action~$W$ non-unique at following points.  Because the MPEP is
initially assumed to lie along the $x$-axis, we study the validity of the
WKB approximation by introducing the expansion
\begin{equation}
W(x,y)=f_0(x)+f_2(x)y^2+o(y^2).
\end{equation}
Physically, $f_2(x)^{-1/2}\epsilon^{1/2}$~is the transverse lengthscale
within which $\rho_1(x,y)$ is nonnegligible.  $f_2(x)$~satisfies a
nonlinear Riccati equation, described in Refs.~\cite{MS1,MS2}.

We begin by considering the drift field of~(\ref{eq:drift}) when $\mu=1$.
For~all $\alpha<4$, we~find numerically that $f_2$~remains positive and
finite from $x=1$ to $x=0$, converging to~$1$ as the origin is approached.
(Because the origin is a fixed point of~${\bf u}$, it~is formally reached in
infinite time; see the discussion in Ref.~\cite{MS2}).  However when
$\alpha=4$, $f_2$~converges to {\em zero\/}, signalling the appearance of a
transverse `soft~mode,' or instability.  (This is accompanied by a
divergence of the prefactor $K$ in~(\ref{eq:WKB}).)  For~all $\alpha>4$,
$f_2$~reaches zero at {\em positive\/} $x$, and then plunges to~$-\infty$
in {\em finite\/} time (hence at strictly positive~$x$).  This signals the
presence of a singularity in~$W$.

Fig.~\ref{fig:curves}(a) displays this graphically.  As~one moves from~$(1,0)$
toward the origin, the iso-action curves become non-convex
at the point where $f_2$~goes negative.  They become increasingly pinched
near the $x$-axis, leading to a cusp singularity at~$(x_F,0)$, where
$f_2$~diverges.  Beyond this point $W$~remains continuous, but is not
continuously differentiable across the $x$-axis.  It~must be computed by
minimizing over trajectories, and different trajectories are employed
as~$y\to0^+$ and~$y\to0^-$.

Fig.~\ref{fig:curves}(b) makes this clear; it~shows the zero-energy
classical trajectories, derived from the Wentzell-Freidlin Hamiltonian by
integrating Hamilton's equations.  As~$x$ is decreased from~$1$, off-axis
trajectories eventually reconverge on the axis, signalling that $(x_F,0)$
is a focus.  (It~is not hard to show this analytically, but we omit the
proof here.)  Every point~$(x,0)$ with~$x<x_F$, including the saddle point
at the origin, is reachable by {\em three\/} zero-energy trajectories: the
original $x$-axis MPEP, and two symmetrically placed off-axis trajectories,
with lesser (degenerate) actions.

The dependence of the focus~$x_F$ on~$\alpha$, for the particular drift
field~(\ref{eq:drift}), is of some interest.  When~$\alpha=4$, $x_F=0$.
As~$\alpha$ increases beyond~$4$, $x_F$~increases continuously from zero
but never reaches~$x_S$.  When~$\alpha =9$, a~new focus appears at the
origin so that {\em two\/} foci occur on the $x$-axis.  Both of these move
to larger~$x$ as $\alpha$~increases.  We~have found numerically that a new
focus appears at~$\alpha=n^2$,\ $n=2,3,4,\ldots$, but have not yet explored
in~detail the set of zero-energy trajectories when multiple foci occur.
We~restrict ourselves below to the case of a single focus,
{\em i.e.}, if~$\mu=1$ we~consider only $4<\alpha<9$.
If~$\mu\neq1$ the range of allowed values of~$\alpha$ will differ.

Upon encountering a single focus, the MPEP {\em bifurcates\/}.  The path of
minimum classical action --- hence, the MPEP --- to any point~$(x,0)$
with~$x>x_F$ from~$(x_S,0)$ remains the $x$-axis.  For~$x<x_F$, however,
the former MPEP ({\em i.e.},~the $x$-axis) remains
an extremum of the action but
is no longer a local minimum~\cite{Schulman}.  There are two new MPEPs,
related by $y\to-y$.  The symmetry has been broken: the drift field and the
equations of motion are symmetric about the $x$-axis, but each MPEP is~not.

The new (curved) MPEPs, and their common action, are easily computed
numerically.  Since their action is the exponential growth rate of the
MFPT, the bifurcation greatly affects the low-noise asymptotic behavior.
In Fig.~\ref{fig:actiongraph} we plot the classical action of the MPEP(s)
between the stable and hyperbolic points (and therefore the asymptotic
slope of the MFPT {\em vs.}~$1/\epsilon$ on a log-log plot) {\em
vs.}~$\alpha$ for the case $\mu=1$ in the drift field~(\ref{eq:drift}).
The onset of focusing at~$\alpha=4$ is evident.  The MFPT prefactor, one
can show, is also singular at~$\alpha=4$.

To~study the effects of focusing on the exit location distribution near the
saddle, we~set up a covariant formalism for computing the WKB behavior
of~$\rho_1$ along a curved MPEP.  The Hamilton-Jacobi equation is
\begin{equation}
0=H(x^i,p_i)={\textstyle{\frac12}}D^{ij}p_ip_j+u^i({\bf x})p_i,
\end{equation}
where
$p_i=\partial W/\partial x^i$, an index is summed over when repeated, and
the diffusion tensor~${\bf D}$ may be anisotropic (but~we consider only
isotropic~$\bf{D}$ here).  We~adopt the notation $p_{i,j}=\partial
p_i/\partial x^j$; note that $W,_{ij}=p_{i,j}$.  By~twice differentiating
the Hamilton-Jacobi equation we arrive at the evolution equation for the
second derivatives of~$W$ along any zero-energy trajectory:
\begin{equation}
\label{eq:fluctevolution}
\dot W_{,ij}=-D^{kl}W_{,ki}W_{,lj}-u^k,_iW_{,kj}-u^k,_jW_{,ki}-u^l_{,ij}p_l
\end{equation}
where the dot signifies a derivative with respect to transit time along the
trajectory.

Eq.~(\ref{eq:fluctevolution}) describes the evolution of fluctuations in
all directions of the action along the MPEP, and represents a
generalization of our previously derived nonlinear Riccati
equation~\cite{MS1,MS2} (to~which it~reduces when $i=j=y$ and the MPEP is
along the $x$-axis).  Similarly, the covariant generalization of our
transport equation for~$K$ is~\cite{Talkner87}
\begin{equation}
\label{eq:Keqn}
\dot K/K=-u^i_{,i}-{\textstyle{\frac12}}D^{ij}W_{,ij}.
\end{equation}
It~is amusing to note that the diffusion tensor plays a role similar to
that of a metric tensor, with additive noise (constant~${\bf D}$)
corresponding to flat space, while multiplicative noise introduces an
effective curvature~\cite{Graham77}.

In the vicinity of the stable and hyperbolic fixed points
($t\to-\infty,\infty$~above) the momentum and therefore the inhomogeneous
term of~(\ref{eq:fluctevolution}) vanish.  We~see that in these limits we
must solve an algebraic Riccati equation
\begin{equation}
\label{eq:algebraic}
D^{kl}W_{,ki}W_{,lj}+u^k,_iW_{,kj}+u^k,_jW_{,ki}=0.
\end{equation}
Eq.~(\ref{eq:algebraic}) was also derived by Ludwig and Mangel in the
context of stochastic ecology~\cite{Ludwig75}.
%
The quasistationary density
in the far field of the `diffusive' lengthscale, of size
$O(\epsilon^{1/2})$, surrounding each fixed point is accordingly (up~to a
constant factor)
\begin{equation}
\label{eq:Gaussian}
\rho_1({\bf x})\sim\exp
\left[-({\bf x}-{{\bf x}}_0){{\bf C}}^{-1}({\bf x}-{\bf x}_0)/2\epsilon\right]
\end{equation}
where ${\bf x}_0=(x_S,0)$ or~$(0,0)$, and the inverse covariance matrix
${\bf C}^{-1}=(C^{-1})_{ij}=W_{,ij}$ is determined by the matrix
equation~(\ref{eq:algebraic}).  ${\bf C}^{-1}$~will be positive definite
at the stable point, but not at the saddle point.

Our procedure is to compute $W({\bf x})$ and $K({\bf x})$ by (numerically)
integrating the coupled equations (\ref{eq:fluctevolution})
and~(\ref{eq:Keqn}) along the curved MPEP.  Close to the two fixed points,
we~match to a quasistationary density which (a)~is a solution of the
Fokker-Planck equation, and (b)~matches with~(\ref{eq:Gaussian}) on~the
$O(\epsilon^{1/2})$ lengthscale.

We find that the asymptotic behavior of exit phenomena falls into three
qualitatively different categories depending on whether the parameter~$\mu$
in the drift field is greater than, less than, or equal to~$1$.  This is
because in the vicinity of the saddle at the origin, ${\bf u}\approx(x,-\mu
y)$.  In~this linearized field Hamilton's equations are easily solved.
(All trajectories possess a reflection symmetry about the $x$-axis; when we
refer to a trajectory in the first quadrant, note that it has a mirror
image in the fourth.)  The zero-energy trajectory which approaches the
origin from above follows a curve $y\propto x^\mu$; hence the MPEP
asymptotically approaches the $x$-axis for~$\mu>1$, the $y$-axis
for~$\mu<1$, and comes in at a fixed angle for~$\mu=1$.
[Cf.~Fig.~\ref{fig:curves}(b).]  We~consider these cases in~turn.

$\mu>1$: Very near the saddle point at~$(0,0)$, this case resembles the
case of unbroken symmetry described in Ref.~\cite{MS2}.  The action there
can be approximated as $W(x,y)\approx -x^2+\mu y^2$.  The prefactor~$K$
simply tends to a nonzero constant.  Because the MPEPs approach the origin
asymptotically close to the $x$-axis, matching in the vicinity of the
saddle remains the same as in the unbroken case.  Up~to a constant factor,
the quasistationary density~$\rho_1$ on the $x,y=O(\epsilon^{1/2})$
lengthscale will be
\begin{equation}
\left[
\exp (x^2/2\epsilon)\,
y_2\bigl({\textstyle{1\over2}},x\sqrt{2/\epsilon}\bigr)\right]
\exp (-\mu y^2/\epsilon)
\end{equation}
where $y_2(\frac12,\cdot)$ is the odd parabolic cylinder
function~\cite{Abramowitz65} of index~$\frac12$.  This agrees
with~(\ref{eq:Gaussian}) in the far field, and falls to zero as~$x\to0$ due
to absorption of probability.  The exit location distribution will be
asymptotically Gaussian, with standard deviation
$\epsilon^{1/2}/\sqrt{2\mu}$, as~in the unbroken case.  We~stress, however,
that in this and the following two cases the exponential dependence of the
MFPT on the noise depends on proper identification of the
$\alpha$-dependent off-axis MPEPs, as does the prefactor.

$\mu<1$: Here the MPEP approaches
the origin asymptotically along the $y$-axis.
However, the {\em momentum\/} of the MPEP asymptotically aligns
itself along the $x$-axis.  (This is true in all three cases.)  This can
easily be seen by solving Hamilton's equations of motion near the origin.
Because the quadratic fluctuations of the action around the MPEP are
perpendicular to the momentum, we~have the unusual situation where
`transverse' fluctuations of the escaping particle around the MPEP become
asymptotically aligned with the MPEP itself: they become longitudinal.
The quasistationary density and exit location distribution are, moreover,
spread over an anomalously large region.  Absorption takes place in a
boundary layer of width~$O(\epsilon^{1/2})$, but because the MPEP
approaches the origin along a curve $y\propto x^\mu$ the appropriate
lengthscale in the $y$-direction is $y=O(\epsilon^{\mu/2})$,
not~$y=O(\epsilon^{1/2})$.

This becomes clear when $W(x,y)$ and~$K(x,y)$ are approximated near the
origin.  Integrating the evolution equation~(\ref{eq:fluctevolution}),
in~the linear approximation, yields
\begin{equation}
\label{eq:Wskewed}
W(x,y)\approx -2x(|y|/A)^{1/\mu} + (|y|/A)^{2/\mu}
\end{equation}
where $A$ is the limit of the ratio $y/x^{\mu}$ along the MPEP as it
approaches the origin.  Since $\mu<1$, this action is of higher order than
quadratic in the displacements~$x,y$.  Equivalently, the inverse covariance
matrix~${\bf C}^{-1}$ of~(\ref{eq:Gaussian}) {\em vanishes\/}.  This
vanishing is always allowed [$W_{,ij}\equiv0$ is a solution of the
algebraic Riccati equation~(\ref{eq:algebraic})] but it occurs only
when~$\mu<1$, leading to an anomalously large lengthscale.  The behavior of
the prefactor~$K$ may be computed from~(\ref{eq:Keqn}); along the MPEP,
it~is asymptotically proportional to~$x^{1-\mu}$.  This is an unusual
situation: strictly speaking,
the `frequency factor'~\cite{MS2,Landauer61}
$K(0,0)$~equals zero, yet all physical exit quantities remain nonzero and
finite.  The MFPT prefactor, however, turns out not to be anomalous; it~is
independent of~$\epsilon$.

To compute the exit location density and the MFPT asymptotics, an
expression for the quasistationary density~$\rho_1$ on the
$x=O(\epsilon^{1/2})$, $y=O(\epsilon^{\mu/2})$ lengthscale is needed.
It~is easily seen to be (up~to a constant factor)
\begin{displaymath}
|y|^{(1/\mu)-1}
\sinh[2x(|y|/A)^{1/\mu}/\epsilon]\,\exp[-(|y|/A)^{2/\mu}/\epsilon]
\end{displaymath}
which both solves the Fokker-Planck equation to leading order, and matches
up with the WKB approximation~(\ref{eq:WKB}) when $W$~is given
by~(\ref{eq:Wskewed}) and $K\sim x^{1-\mu}$ along the MPEP.  Since the exit
location probability density is proportional to~$\partial_n\rho_1$,
{\em i.e.},
$\partial\rho_1/\partial x(x=0)$, this gives an asymptotic exit location
distribution $f(y)\,dy$ on~the $y$-axis, with
\begin{equation}
f(y)\propto |y|^{(2/\mu)-1} \exp[-(|y|/A)^{2/\mu}/\epsilon]
\end{equation}
So on the $O(\epsilon^{\mu/2})$ lengthscale to either side of the saddle,
the exit location has a bimodal (symmetrized) Weibull
distribution~\cite{Burington70}, with shape parameter~$2/\mu$.
(Equivalently, $|y|^{2/\mu}$ at~exit time becomes {\em exponentially
distributed\/} in the low-noise limit.)  This `skewing' phenomenon (the
particle tending to exit to one side of a saddle, in a non-Gaussian way,
rather than at the saddle itself) was discovered by Bobrovsky and
Schuss~\cite{Bobrovsky82}, and has been further
investigated~\cite{Bobrovsky92,Day93}.  However, in~our simple model we are
able to work out the skewed exit location distribution exactly.  Of~course
the scaling parameter~$A=A(\alpha,\mu)$ must be determined numerically,
by~integrating Hamilton's equations.  The physical interpretation of the
exponential distribution remains unclear.

$\mu=1$: As in the $\mu>1$ case, absorption occurs on the
$x,y=O(\epsilon^{1/2})$ lengthscale.  But a new feature enters: the inverse
covariance matrix~${\bf C}^{-1}$ is no~longer uniquely determined
by~(\ref{eq:algebraic}).  There is a free parameter, which must be computed
numerically by matching.  The asymptotic expression for the exit location
distribution involves parabolic cylinder functions.  We~defer the details.

We~have dealt only with a drift field symmetric about some axis.  Our
results (excepting those relating to broken symmetry) and methods are more
general and apply to drift fields without such symmetry.  So~they should
apply, in~particular, to the asymmetric models (with
anisotropic~$\bf{D}$) reviewed by Bobrovsky and
Zeitouni~\cite{Bobrovsky92}.  For such models the MPEP will typically be
curved even in the absence of focusing, and skewed exit distributions will
frequently occur.  But the presence of a focus will still be signalled by a
breakdown in transverse fluctuations, at~which point qualitatively new
behavior will emerge.  One amusing prospect is to study drift fields with a
`symmetry parameter.' As~this is tuned through zero one might observe a
jump in the exponential growth rate of the MFPT, much as a jump in
magnetization results when an applied field passes through zero.



This research was partially supported by the National Science Foundation
under grant NCR-90-16211 (RSM), and by the U.S. Department of Energy under
grant DE-FG03-93ER25155 (DLS).

\begin{figure}
\caption{(a)~The curves of constant action~$W$ surrounding the stable fixed
point $(x_S,0)=(1,0)$, revealing the presence of a singularity in~$W$ at
$(x_F,0)\approx(0.3,0)$. Here $\mu=1$ and $\alpha=5$. (b)~The classical
trajectories of zero energy, emanating from $(1,0)$, from which the
iso-action curves are computed.  The two MPEPs are the off-axis trajectories
incident on~$(0,0)$.\label{fig:curves}}
\end{figure}

\begin{figure}
\caption{The action of the MPEP(s) for
the drift field~(\protect\ref{eq:drift}),
as a function of the parameter~$\alpha$ when~$\mu=1$. The MPEP extends
along the $x$-axis only if ${\alpha\le4}$, and no focus is
present.\label{fig:actiongraph}}
\end{figure}

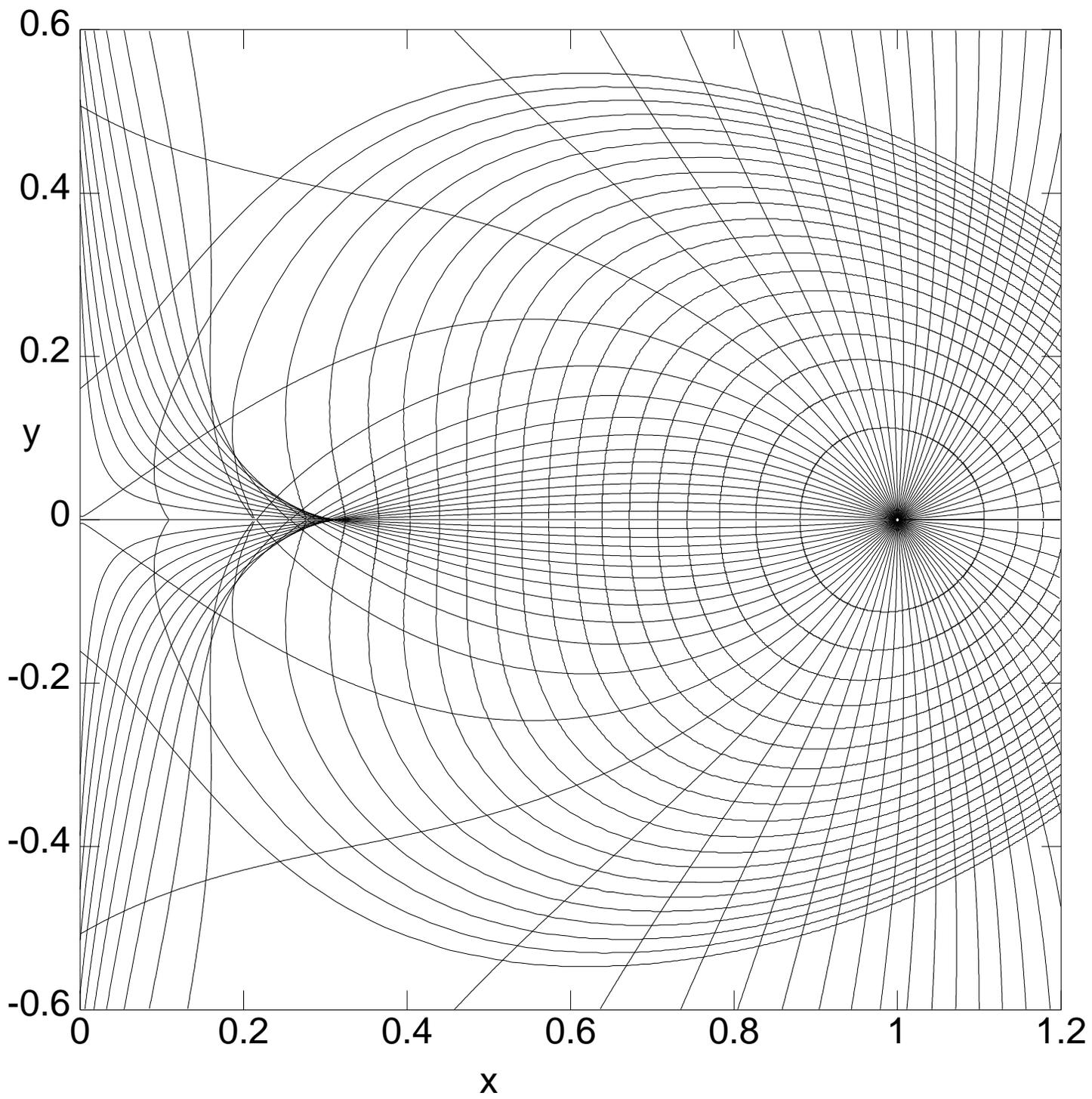
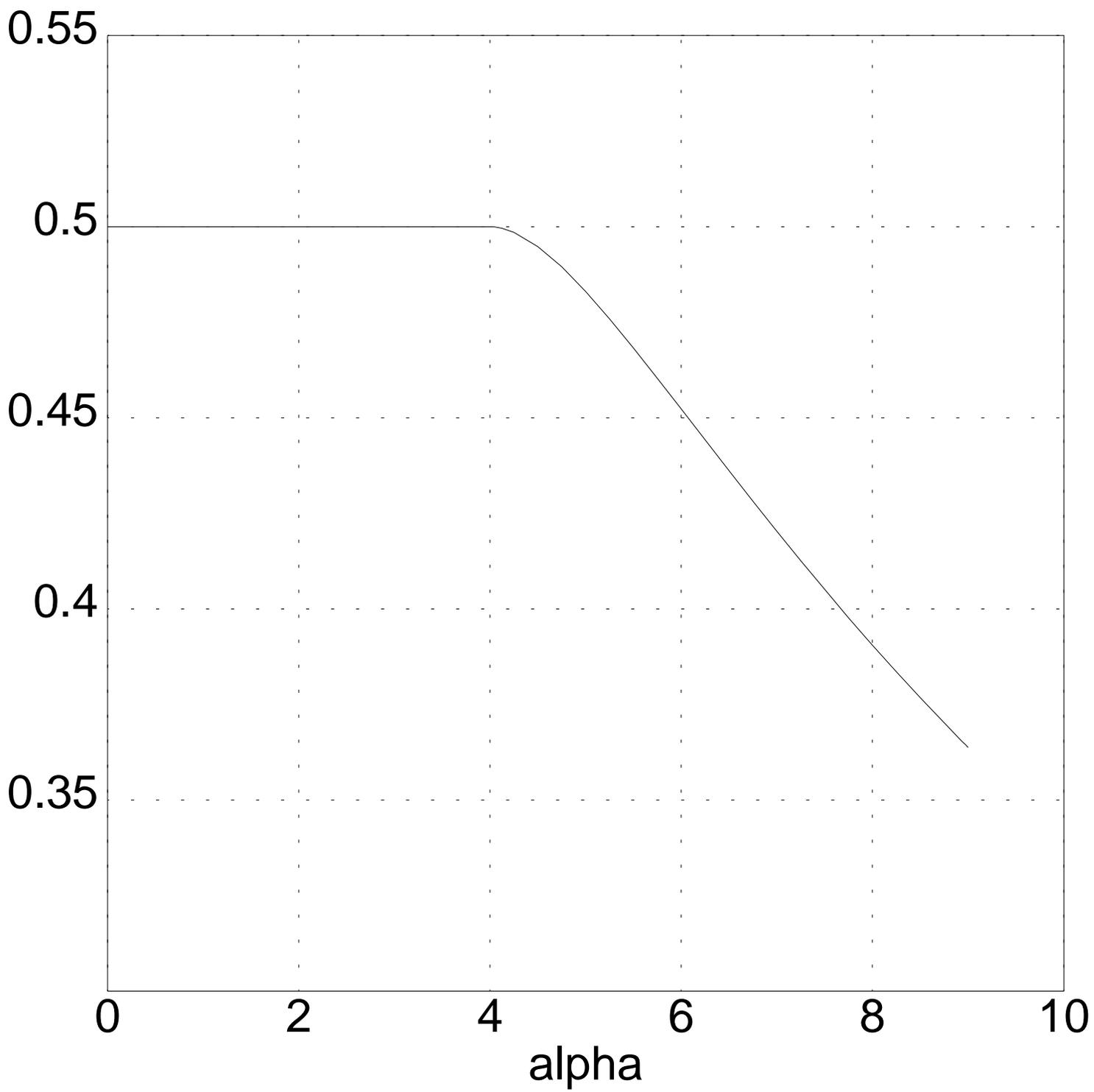
\end{document}